\documentclass{vgtc}                          

\ifpdf
  \pdfoutput=1\relax                   
  \usepackage{graphicx}                
  \DeclareGraphicsExtensions{.pdf,.png,.jpg,.jpeg} 
\else
  \usepackage{graphicx}                
  \DeclareGraphicsExtensions{.eps}     
\fi%

\graphicspath{{figures/}{pictures/}{images/}{./}} 

\usepackage{microtype}                 
\PassOptionsToPackage{warn}{textcomp}  
\usepackage{textcomp}                  
\usepackage{mathptmx}                  
\usepackage{times}                     
\usepackage{cite}                      
\usepackage{tabu}                      
\usepackage{booktabs}                  
\usepackage{xspace}
\usepackage{url}
\usepackage{hyperref}
\usepackage{wrapfig}
\usepackage{amsmath}

\onlineid{0}

\vgtccategory{Research}

\vgtcinsertpkg

\newcommand{\system}{\textsc{ScatterSearch}\xspace}
\newcommand{\stitle}[1]{\noindent\textbf{#1}}
\newcommand{\npar}{\par\noindent}

\title{\system: Visual Querying of Scatterplot Visualizations}

\author{Doris Jung-Lin Lee\\ %
    \scriptsize dorislee@berkeley.edu\\
    \scriptsize UC Berkeley
\and Jaewoo Kim \quad Renxuan Wang\\
    \scriptsize \{jkim475,renxuan2\}@illinois.edu\\
     \scriptsize University of Illinois, Urbana-Champaign
\and Aditya Parameswaran\\ %
    \scriptsize adityagp@berkeley.edu\\
     \scriptsize UC Berkeley
}

\abstract{
    Scatterplots are one of the simplest and most commonly-used visualizations for understanding quantitative, multidimensional data. However, since scatterplots only depict two attributes at a time, analysts often need to manually generate and inspect large numbers of scatterplots to make sense of large datasets with many attributes. We present a visual query system for scatterplots, \system, that enables users to visually search and browse through large collections of scatterplots. Users can query for other visualizations based on a region of interest or find other scatterplots that ``look similar'' to a selected one. We present two demo scenarios, provide a system overview of \system, and outline future directions. 
} 

\nocopyrightspace

\begin{document}


\firstsection{Introduction}
\maketitle
\par Scatterplots are simple yet powerful visualizations for highlighting relationships between quantitative variables (commonly referred to as \emph{measures}). Despite the success of scatterplots, there is a growing trend towards ``wide'' datasets with many measures, due to the ease of data acquisition in a variety of domains. This is certainly true in material science, where large numbers of physical measurements are recorded for different solvents~\cite{Lee2019}, with the number of potential scatterplots growing quadratically with the number of attributes in the dataset. Given these large collections, it is often impossible for an analyst to know apriori which variables would lead to visualizations of interest, without comparing across all possible scatterplots that could be generated from the multidimensional dataset.
\par Existing multidimensional data exploration methods include scatterplot matrices (SPLOM), dimensionality reduction, and high-dimensional data visualization techniques such as parallel coordinates, but they often suffer from overplotting, resulting in loss of information and insights. To this end, several existing systems have been developed for interactive exploration of multi-dimensional data through scatterplots, including navigation through the scatterplot matrix~\cite{Elmqvist2008} and exploration of the space of graph-theoretic features (scagnostics)~\cite{Dang2014} or statistical features~\cite{Seo2004} of the visualizations. However, these techniques either still require users to manually navigate through large numbers of visualizations or are based on metrics that do not reliably capture human similarity perception of scatterplots~\cite{Pandey2016}. Our goal in this work is to let users directly search for scatterplots by specifying their desired pattern characteristics. 
\par Systems that enable users to visually specify and search for visualizations are known as \emph{visual query systems} (VQSs). VQSs are known for their responsive and intuitive interfaces, which allows novices to search for visualizations with via flexible querying capabilities~\cite{Lee2019,Correll2016,Mannino2018,ryall2005querylines,Hochheiser2004}.
However, these systems have largely been focused on visual querying for line charts, rather than scatterplots. Unlike line charts, scatterplots are two-dimensional rather than one-dimensional, making it harder to represent and compare against a large collection. In particular: 
\par \stitle{Challenge \#1: Specification Interface for Pattern Queries}
\npar To query for line charts, many systems employ query-by-sketching to enable users to specify a sequence of datapoints as the desired pattern of interest. However, a sketched 1-D line does not completely capture the distributional aspects of a scatterplot, such as density or variance in particular regions. The question of how to design an interface to elicit pattern information from users is an unsolved problem. We introduce two query specification techniques in this paper: {\em region queries} and {\em query-by-scatterplot} and describe potential extensions to this work.
\par \stitle{Challenge \#2: Perceptual Similarity for Scatterplot Matching} 
\npar To find a visualization that matches a desired pattern, visual query systems rank a collection of potentially-matching visualizations using a scoring function. The scoring function is often a distance metric that defines whether a target visualization ``looks similar'' to a specified visualization. While similarity metrics for 1D line charts is a well-studied problem in data mining, the notion of perceptual similarity between 2D distributions (scatterplots) is not as well-defined. For image retrieval, Earth mover's distance (EMD) is a common metric for characterizing the amount of work it takes to transform one distribution into another distribution~\cite{Rubner2000}. However, computing EMD in practice can be computationally expensive, as there is no closed-form solution to the optimization problem. In this work, we propose a novel multi-level distance metric (associated with a preprocessing pipeline) intended to capture different perceptual aspects of scatterplots that can be computed at interactive speeds.
\section{Interaction/Demo Scenario}
We first present two usage scenarios to demonstrate the two different classes of queries supported in \system, then we describe the system in more detail in the next section.
\begin{figure}[h!]
    \vspace{-8pt}
    \centering
    \includegraphics[width=0.85\linewidth]{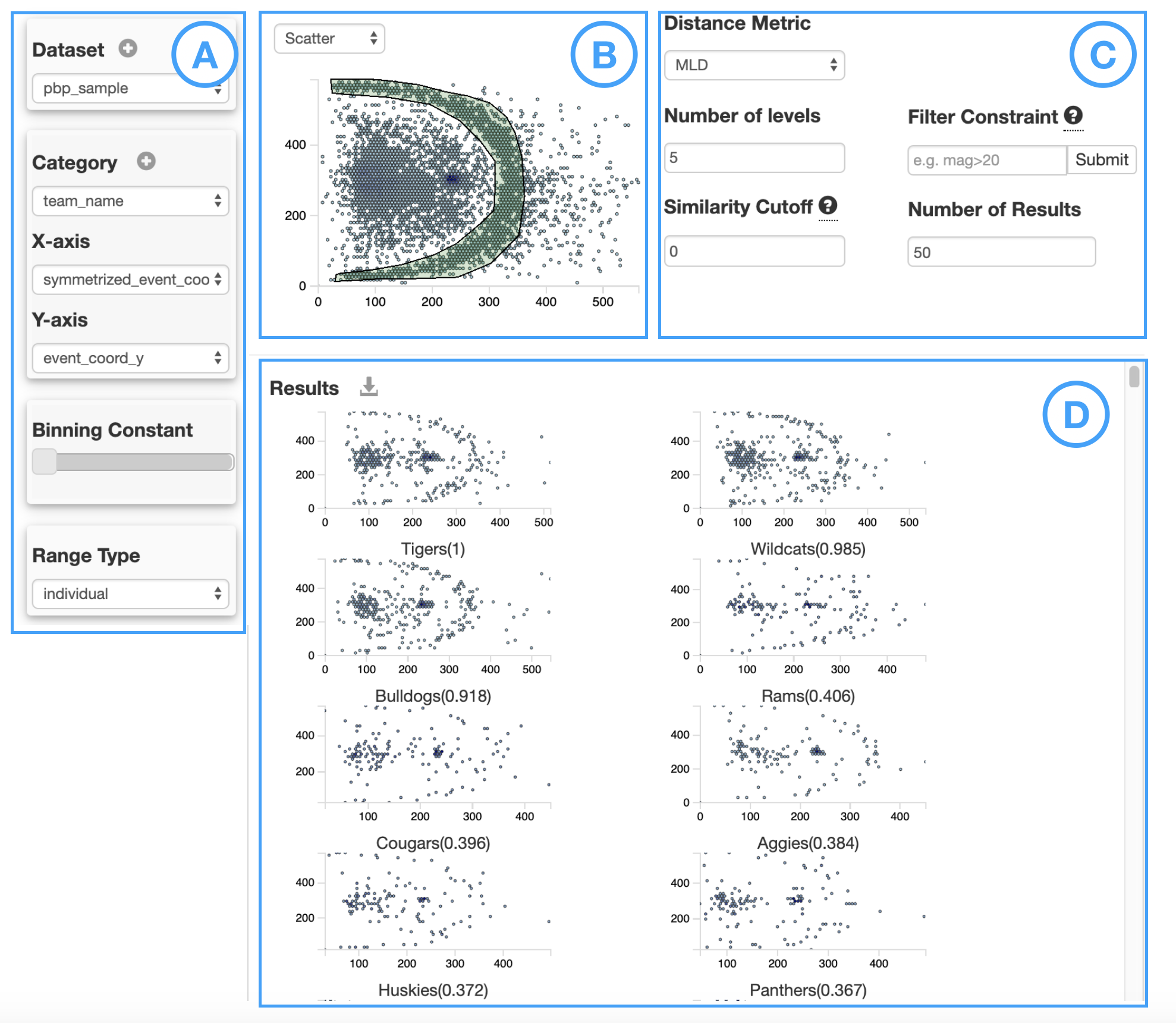}
    \vspace{-5pt}
    \caption{Overview of \system interface.}
    \label{fig:interface}
    \vspace{-8pt}
\end{figure}
\\\stitle{Scenario \#1: Region Query for Sports Analytics}
\par\noindent Jane is a sports analyst interested in how different plays and strategies are used by NCAA basketball teams in different parts of the basketball court. She loads in the NCAA play-by-play data\footnote{{\tiny\url{console.cloud.google.com/marketplace/details/ncaa-bb-public/ncaa-basketball}}} into \system and examines the player's positions on the court, broken down by the different event types (\texttt{event\_type}). She sees an overview of all the datapoints plotted on the querying canvas and notices that there is a dense region near the free-throw line. She draws a region query on the canvas (Figure~\ref{fig:interface}B green region) and found that many of these events were indeed free throws, but there were also rebounds that happened near the free-throw line. To look at shots that were made in the game, she switches the category to \texttt{shot\_type}. She finds it difficult to compare the relative locations of the different shot types on the court, so she selects the option to fix the x and y extent of the resulting visualizations. She learns that while dunks, layups, hook shots, and tip shots happen in close proximity to the hoop, jump shots can pretty much happen anywhere on the court and account for most of the three-pointers. She is now interested in finding teams that make a lot of three-pointers. She selects the region around the three-point line and finds that the Tigers, Wildcats, and Bulldogs have many plays around the three-point line.
\\ \textbf{Scenario \#2: Query-by-Visualization for Social Data Analysis}
John is a social scientist interested in relationships between different socio-economic factors and their connection with crime rates across different communities. He examines a sampled dataset of 190 visualizations generated from the pairwise combination of 20 different quantitative attributes in the dataset\footnote{{\tiny\url{archive.ics.uci.edu/ml/datasets/communities+and+crime}}}. By first browsing through the data, he finds that the percentage of households with rent and investment incomes is inversely correlated with the percentage of households that are on public assistance. He wants to find other pairwise attributes that also have a similar trend, so he drags and drops this visualization onto the canvas (Figure~\ref{fig:interface}B) as a query. He finds several variables that are inversely correlated with elderly, retirement communities, such as lower percentages of the population between ages 12-29 (\texttt{agePct12t29}) and smaller household size (\texttt{householdsize}). He also finds that as the family median income in a community increases, its need for public assistance decreases.
\section{System Description}
\setlength{\columnsep}{5pt}%
\begin{wrapfigure}{r}{0.5\linewidth}
\vspace{-15pt}
    \centering
    \includegraphics[width=\linewidth]{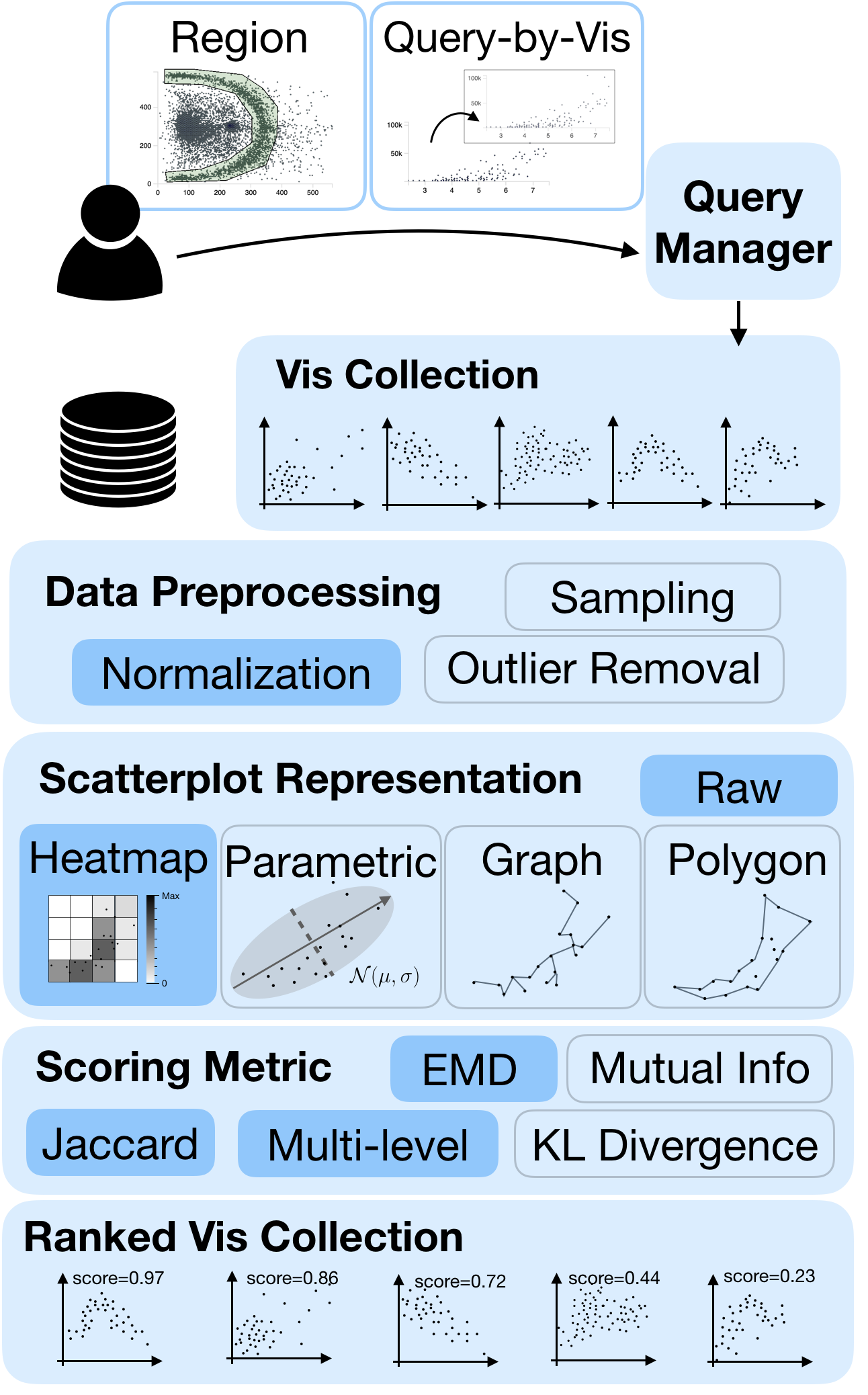}
    \caption{\system components. Blue boxes represent existing features, grey boxes are extensible modules that can be implemented in future work.}
    \label{fig:architecture}
    \hspace{-20pt}
\vspace{-15pt}
\end{wrapfigure}
Figure~\ref{fig:architecture} provide an overview of the \system system components. \system is built as an extension to Zenvisage~\cite{Lee2019}, a VQS for line charts. Users can either issue a region query or query by dragging and dropping an existing visualization. Based on the selected axes, the query manager generates a collection of scatterplot visualizations by retrieving the data values from the underlying database. Each scatterplot consists of a series of X, Y datapoints. These raw values can be preprocessed in a way that improves the downstream tasks. For example, normalization and outlier removal accentuates the key patterns in the scatterplot and sampling decreases the amount of data that needs to be processed. Next, each scatterplot is further processed into some representation that facilitates effective similarity computation with the queried scatterplots. For example, a scatterplot can either be binned into a heatmap matrix, parametrized into a functional representation, or represented as a graph or polygon. In the current version of \system, we use the raw representation for region queries and heatmap representation for query-by-visualization. The scatterplot representations are inputs to the scoring functions, which produce a score indicating how well a scatterplot satisfies the query. Finally, the scatterplots are sorted based on this score and displayed as a ranked list to the user.
\newpage
\stitle{Scoring Metrics:}
For query-by-visualization, we developed a novel multi-level Euclidean distance that operates on the heatmap matrix representations of the scatterplot. To compute the similarity between two matrices $M_a$ and $M_b$, we first generate a set of matrix representations of various resolutions for each scatterplot (e.g. $M_a = \{M_{2x2},M_{4x4},...,M_{1024x1024}\}$). We compute the similarity between the two scatterplots based on comparing their corresponding grids at the same resolution. The resulting distances are aggregated into a single score based on a weighted sum (with constants $k_i$).
\begin{equation*}
MLD(M_a,M_b) = k_1\cdot D(M_{2x2}^a,M_{2x2}^b)+k_2\cdot D(M_{4x4}^a,M_{4x4}^b)+...
\end{equation*}
The intuition for computing a score across multiple resolutions is that coarse-grained grids give a general sense of the scatterplot density at different locations and are cheaper to compute than fine-grained grids. Accordingly, we assign higher weights ($k_i$) to coarse-grained grids since if a pair of scatterplots are not ``roughly'' similar on a coarse-grained grid, it is most likely not very similar at the finer level and overall. For the region query, we compute the number of points that lie inside the selected region as the score. 
\section{Conclusion \& Future Work}
\par We presented \system, a novel VQS for scatterplot visualizations. To our knowledge, \system is the first end-to-end system that enables users to visually query for desired scatterplots of interest. Our preliminary prototype serves as an experimentation framework to evaluate the effect of different query specification interfaces, preprocessing procedures, representations, and similarity metrics. With this modular framework, our next step is to work with real-world users and datasets to evaluate the efficacy of \system's pattern matching capabilities, as well as understanding the types of queries that users are interested in when searching for scatterplots. We hope to gather formative feedback from the VIS community through the poster demonstration to improve this work.
\vspace{-5pt}
\bibliographystyle{abbrv}
{\footnotesize\bibliography{main}}
\end{document}